\documentclass[useAMS,usenatbib]{mn2e}
\usepackage{myaasmacros,epsfig}

\def\ltsima{$\; \buildrel < \over \sim \;$}
\def\simlt{\lower.5ex\hbox{\ltsima}}   
\def\gtsima{$\; \buildrel > \over \sim \;$}
\def\simgt{\lower.5ex\hbox{\gtsima}}
\newcommand\bcite[1]{\citeauthor{#1} \citeyear{#1}}

\title[The importance of tides for dSphs]
{The importance of tides for the Local Group dwarf spheroidals}

\author[Read et. al.]{J. I. Read $^1$\thanks{Email: jir22@ast.cam.ac.uk} 
M. I. Wilkinson $^1$ N. Wyn Evans $^1$ G. Gilmore $^1$ \& Jan T. Kleyna $^2$ 
\\ $^1$Institute of Astronomy, Cambridge University, Madingley Road, 
Cambridge, CB3 OHA, England
\\ $^2$Institute for Astronomy, University of Hawaii, 2680 Woodlawn Drive, Honolulu, HI 96822}

\begin{document}

\maketitle

\begin{abstract}
There are two main tidal effects which can act on the Local Group
dwarf spheroidals (dSphs): tidal stripping and tidal shocking. Using
N-body simulations, we show that tidal stripping always leads to flat
or rising projected velocity dispersions beyond a critical radius; it
is $\sim 5$ times more likely, when averaging over all possible
projection angles, that the cylindrically averaged
projected dispersion will rise, rather than be flat. In
contrast, the Local Group dSphs, as a class, show flat or falling
projected velocity dispersions interior to $\sim 1$\,kpc. This argues for
tidal stripping being unimportant interior to $\sim 1$\,kpc for most
of the Local Group dSphs observed so far. We show that tidal shocking
may still be important, however, even when tidal stripping is
not. This could explain the observed correlation for the Local Group
dSphs between central surface brightness and distance from the nearest
large galaxy.

These results have important implications for the formation of the
dSphs and for cosmology. As a result of the existence of
cold stars at large radii in several dSphs, a tidal origin for the formation
of these Local Group dSphs (in which they contain no dark matter) is strongly
disfavoured. In the cosmological context, a naive solution to the 
missing satellites problem is to allow only the most massive
substructure dark matter halos around the Milky Way to form stars. It
is possible for dSphs to reside within these halos ($\sim
10^{10}$M$_\odot$) and have their velocity dispersions lowered through
the action of tidal shocks, but only if they have a central density
core in their dark matter, rather than a cusp. A central density cusp
persists even after unrealistically extreme tidal shocking and leads
to central velocity dispersions which are too high to be consistent
with data from the Local Group dSphs. dSphs can reside within cuspy
dark matter halos if their halos are less massive ($\sim
10^{9}$M$_\odot$) and therefore have smaller central velocity
dispersions initially.

\end{abstract}


\section{Introduction}\label{sec:introduction}
The Local Group of galaxies provides a unique test-bed for galaxy
formation models and cosmology. The abundance, spatial distribution
and internal mass distribution, of 
satellites within the Local Group can provide sensitive tests of
cosmological predictions (see
e.g. \bcite{1999ApJ...524L..19M}, 
\bcite{1999ApJ...522...82K}, \bcite{2004ApJ...609..482K},
\bcite{2001ApJ...559..754M} and \bcite{2001ApJ...547L.123M}). However,
central to such studies is an understanding of what the Local Group 
satellite galaxies are. Observationally, they are usually split into
three types: dSph galaxies, which typically have old stellar
populations, are spheroidal 
in morphology, lie close to their host galaxy\footnotemark and are devoid of HI
gas; dIrr galaxies, which have younger stellar populations, irregular
morphology, lie further away from their host galaxy and contain
significant HI gas; and the transition galaxies, which are in between
the dSph and dIrr types \citep{1998ARA&A..36..435M}.

\footnotetext{We use the
  terminology `host galaxy' throughout this paper to refer to either the
  Milky Way or M31 depending on which of these is closer to the
  satellite being discussed.}

Since the dSph galaxies lie, in general, closer to their
host galaxy (there are notable exceptions - see
e.g. \bcite{2003AJ....125.1926G}) it has often been argued that the
tidal field of the host galaxy has
an important role to play in the formation and evolution of these
galaxies. \citet{2001ApJ...559..754M} and \citet{2001ApJ...547L.123M}
have suggested that all of the Local Group satellite galaxies started
out looking more like the dIrrs, the dSphs then forming through tidal
transformations. \citet{1989ApJ...341L..41K} and
\citet{1997NewA....2..139K} have argued that the dSphs might have
formed in the tidal tail of a previous major merger; in their model,
dSphs contain no dark matter and are nearly
unbound. \citet{2002MNRAS.335L..84S} have argued that the dSphs
inhabit the most massive substructure dark matter halos, of mass $\sim
10^{10}$M$_\odot$, predicted by cosmological N-body simulations (see
e.g. \bcite{1999ApJ...522...82K} and \bcite{1999ApJ...524L..19M}); in
such a scenario, tides are unlikely to have significantly altered the
visible distribution of stars in these galaxies.

\begin{figure*} 
\begin{center}
\epsfig{file=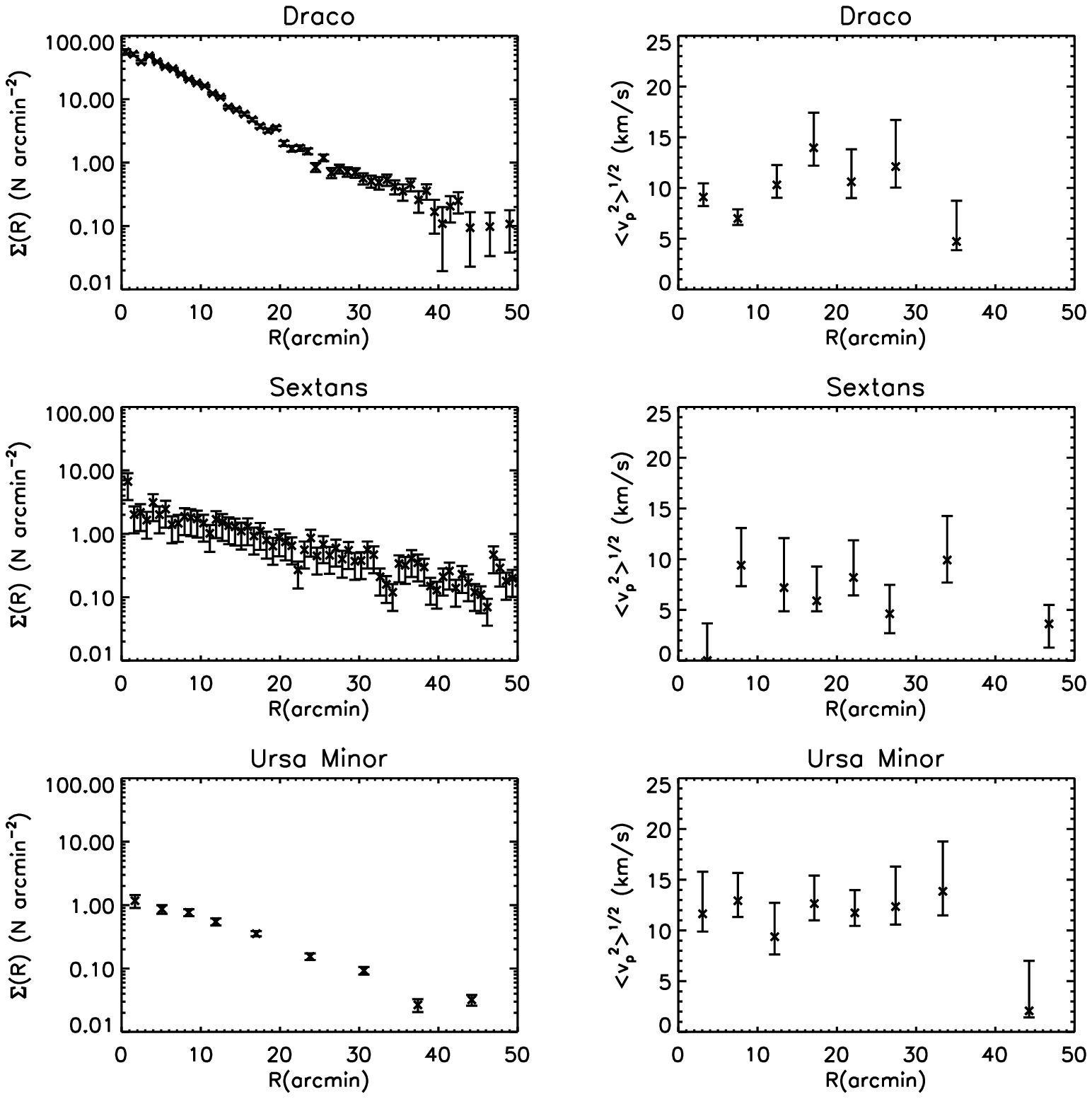,width=16cm}
\caption[]{Surface brightness profiles (left) and projected velocity
  dispersions (right) for three Local Group dSphs observed with good
  stellar kinematics. Data taken from \citet{2004ApJ...611L..21W},
  \citet{2004MNRAS.354L..66K}, \citet{1995MNRAS.277.1354I} and
  \citet{2003AJ....125.1352P}; but see also
  \citet{2001ApJ...549L..63M} for deeper surface brightness data for
  UMi. Notice the similarities between each of the dSphs: the surface
  brightness profiles are roughly exponential, while the projected
  velocity dispersions are approximately flat with a sharp fall-off at
  large radii ($\sim 1$\,kpc).} 
\end{center} 
\label{fig:localdata}
\end{figure*}

In this paper we compare a suite of N-body simulations of dSph
galaxies orbiting in a tidal field, with the latest data from the
Local Group dSphs (see Figure \ref{fig:localdata}). In this way, we
can constrain the importance of tidal effects for these galaxies and
break the degeneracies between the above formation scenarios.

The effects of tides on the Local Group dSphs has been investigated
extensively in the literature, both dynamically (\bcite{1986ApJ...307...97A},
\bcite{1995ApJ...442..142O}, \bcite{1995AJ....109.1071P},
\bcite{2003ApJ...584..541H}, \bcite{2004ApJ...608..663K} and
\bcite{Mashchenko:2005bj}), and
within a cosmological framework  
\citep{2004ApJ...609..482K}. This present work complements that of previous
authors. Like \citet{2003ApJ...584..541H} and
\citet{2004ApJ...608..663K}, we perform simulations which are not
fully cosmologically consistent, but which have the advantage that the
simulated satellites have very high resolution. Unlike these previous
authors, however, our dSph galaxies contain both stars and dark matter. This
allows us to compare our results more easily with the observational
data from the Local Group dSphs.

We consider three possible scenarios for the tidal evolution of the
Local Group dSphs. In the first (model A), we revisit the hypothesis
that the Local Group dSphs contain no dark matter and that their high
velocity dispersions arise instead from the action of
tides\footnote{A dark-matter free model for the formation of the Local
  Group dSphs is often referred to in the literature as the `tidal
  model'. Previous tidal models, such
  as \citet{1997NewA....2..139K}, tried to use tides to explain even
  the central velocity dispersion of the Local Group dSphs. As a
  result, in their models, after a Hubble time, the dSphs became fully
  unbound. In the models we present here, we use unrealistically large
  central mass to light ratios in order to fit the high central
  velocity dispersions of the Local Group dSphs. This means that tides
  need not work quite so hard to reproduce the large outer velocity
  dispersions. If our tidal model is ruled out, more extreme tidal
  models such as that investigated by \citet{1997NewA....2..139K} are also
  ruled out.}. In the
second scenario (models B and C), we 
investigate the hypothesis that the dSphs are dark matter dominated,
but that their outer regions have been shaped by tidal effects. In
this case we ask whether tides could directly cause the drop in
projected velocity dispersion in the outer regions of the dSphs,
recently observed by \citet{2004ApJ...611L..21W}; and see also Figure
\ref{fig:localdata}. In the
third scenario (model D), we investigate a scenario where dSphs were
much more massive in the past. In this case, tidal shocks cause their
central density and velocity dispersions to fall significantly over a
Hubble time, becoming consistent with current data at the present
epoch. It is important to state explicitly here that we do not wish to
model any one specific Local Group dSph, which would require an extensive
search through parameter space and probably not be relevant when taken
out of a cosmological context. Rather, we wish to investigate the generic
effects of tides and how they reshape a dSph. 

This paper is organised as follows: in section \ref{sec:theory}, we
briefly outline some key analytic results for tidal stripping and
tidal shocking - the two dominant tidal effects which act on a
satellite orbiting in a host galaxy potential. The analytic results
and definitions therein will be referred to throughout this paper and
are useful in both understanding and testing 
the simulation results. In section \ref{sec:numerical},
we describe our numerical method for setting up the initial conditions
and integrating the orbit of the satellite in the fixed potential of
the Milky Way. In section
\ref{sec:results}, we present the results from a set of representative
N-body simulations simulations. We demonstrate that tidal stripping
will always lead to a flat or rising projected velocity
dispersion. In section \ref{sec:discussion} we discuss our results in
the context of cosmology; we place mass limits on the Local Group
dSphs. Finally, in section \ref{sec:conclusions} we present our
conclusions.

\section{Theoretical background}\label{sec:theory}

\begin{table*}
\setlength{\arrayrulewidth}{0.5mm}
\begin{center}
\begin{tabular}{lllllllllll}
\hline
{\it Model} & {\it $\rho_*$} & {\it $\rho_{DM}$} & {\it $N_*$} & {\it
  $\xi_*$(kpc)} & {\it $N_{DM}$} & {\it $\xi_{DM}$(kpc)} &
{\it Orbit} & {\it Time(Gyrs)} & {\it $\gamma$}\\
\hline
A & P: 5, 0.23 & None & $10^5$ & $0.01$ & - & - & 9\,kpc,84\,kpc,-34.7$^o$ &
10.43 & 0.0011\\
B & P: 0.072, 0.23 & P: 14, 0.5 & $10^5$ & $0.01$ & $10^6$ & $0.01$ &
23\,kpc,85\,kpc,-34.7$^o$ & 4.5 & 0.00014\\
C & SP: 0.072, 0.1, 0 & SP: 90, 1.956, 1 & $10^5$ & $0.01$ & $10^6$ & $0.01$ & 23\,kpc,85\,kpc,-34.7$^o$ & 4.5 & 0.15 \\
D & SP: 0.072, 0.1, 0 & SP: $10^3$, 1.956, 1 & $10^5$ & $0.01$ &
$10^6$ & $0.01$ & 6.5\,kpc,80\,kpc,7.25$^o$ & 8.85 & 0.83 \\
\hline
\end{tabular}\\
\caption[Initial conditions]{Simulation initial conditions for models
  A-D. The columns from left
to right show the model (labelled by a letter in order of discussion),
the stellar density profile, $\rho_*$, the dark matter density
profile, $\rho_{DM}$, the number of stars, $N_*$, the force softening
for the stars, $\xi_*$, the number of dark
matter particles, $N_{DM}$, the force softening for the dark matter,
$\xi_{DM}$, the dwarf galaxy orbit, the output time in Gyrs at
which we show the results from the model and the strength of tidal
shocks, $\gamma$ (see section \ref{sec:shocking}). The density profiles are
either (P)lummer with parameters: mass ($10^7$M$_\odot$), scale length
(kpc) or (S)plit (P)ower law with parameters: mass
($10^7$M$_\odot$), scale length (kpc) and central log slope, $\alpha$
(see equations \ref{eqn:rhoplum} and \ref{eqn:rhosp} for more
details). The orbit for the dwarf galaxy is given by its pericentre,
apocentre and inclination angle to the Milky Way disc.}
\label{tab:initial}
\end{center}
\end{table*} 

A satellite in orbit around a host galaxy will experience two main
tidal effects which will reshape the stellar and dark matter
distributions: tidal stripping and tidal shocking. Since both of the
mechanisms will be referred to many times throughout this paper, it is
instructive to briefly outline the salient properties of each in this
section.

\subsection{Tidal stripping}\label{sec:stripping}

We discuss the theory of tidal stripping in much more detail in a
companion paper, \citet{Readinprep1}. Here we briefly summarise the
main results of that work. Tidal stripping refers to the capture of
stars by a host galaxy from a satellite galaxy beyond a critical
radius: the tidal radius, $r_t$. In general, the tidal radius
depends upon four factors:  the potential of the host galaxy, the
potential of the satellite, the orbit of the satellite and, which is new to the
calculation presented in \citet{Readinprep1}, {\it the orbit of the
  star within the satellite}. In \citet{Readinprep1}, we
demonstrate that this last point is critical and suggest using {\it
  three tidal radii} to cover the range of orbits of stars within the
satellite. In this way we show explicitly that prograde star orbits
will be more easily stripped than radial orbits; while radial orbits
are more easily stripped than retrograde ones. 

In general, these three tidal radii must be calculated numerically as
outlined in \citet{Readinprep1}. We will use these more accurate
numerical calculations in the presentation of tidal radii
later on in this paper. However, for the special case of
point mass potentials, $r_t$ reduces to the following simple form:

\begin{equation}
r_t \simeq x_p \left(\frac{M_s}{M_g}\right)^{1/3}\left(\frac{1}{1+e}\right)^{1/3}\left(\frac{\sqrt{\alpha^2+1+\frac{2}{1+e}}-\alpha}{1+\frac{2}{1+e}}\right)^{2/3}
\label{eqn:rtperi}
\end{equation}
\noindent
where $x_p$ and $e$ are the pericentre and eccentricity of the
satellite's orbit, $M_s$ is the total mass of the satellite, $M_g$ is
the mass of the host galaxy and $\alpha = -1,0,1$ parameterises the
orbit of the star within the satellite. From equation
\ref{eqn:rtperi}, we can see that stars on prograde orbits ($\alpha=1$) are
more easily stripped than those on radial orbits ($\alpha=0$), which
are more easily stripped than those on retrograde orbits
($\alpha=-1$). Notice that $\alpha=0$ recovers the standard
\citet{1962AJ.....67..471K} tidal radius. Other analytic solutions
also exist for power-law and a restricted class of split power-law
density profiles \citep{Readinprep1}.

Over long times, orbital transformations cause a convergence of these
three tidal radii on the prograde stripping radius and leads to the
onset of tangential anisotropy beyond this point.

A final key result from \citet{Readinprep1} is that the standard
intuition that the tidal radius depends only (within a factor) on the
{\it densities} of the host galaxy and satellite galaxy is only valid
for stars within the satellite which are on radial orbits. For general
star orbits, their stripping radius will depend upon the {\it mass
  distribution} within the satellite and not just the enclosed
mass.

\begin{figure*} 
\begin{center}
\epsfig{file=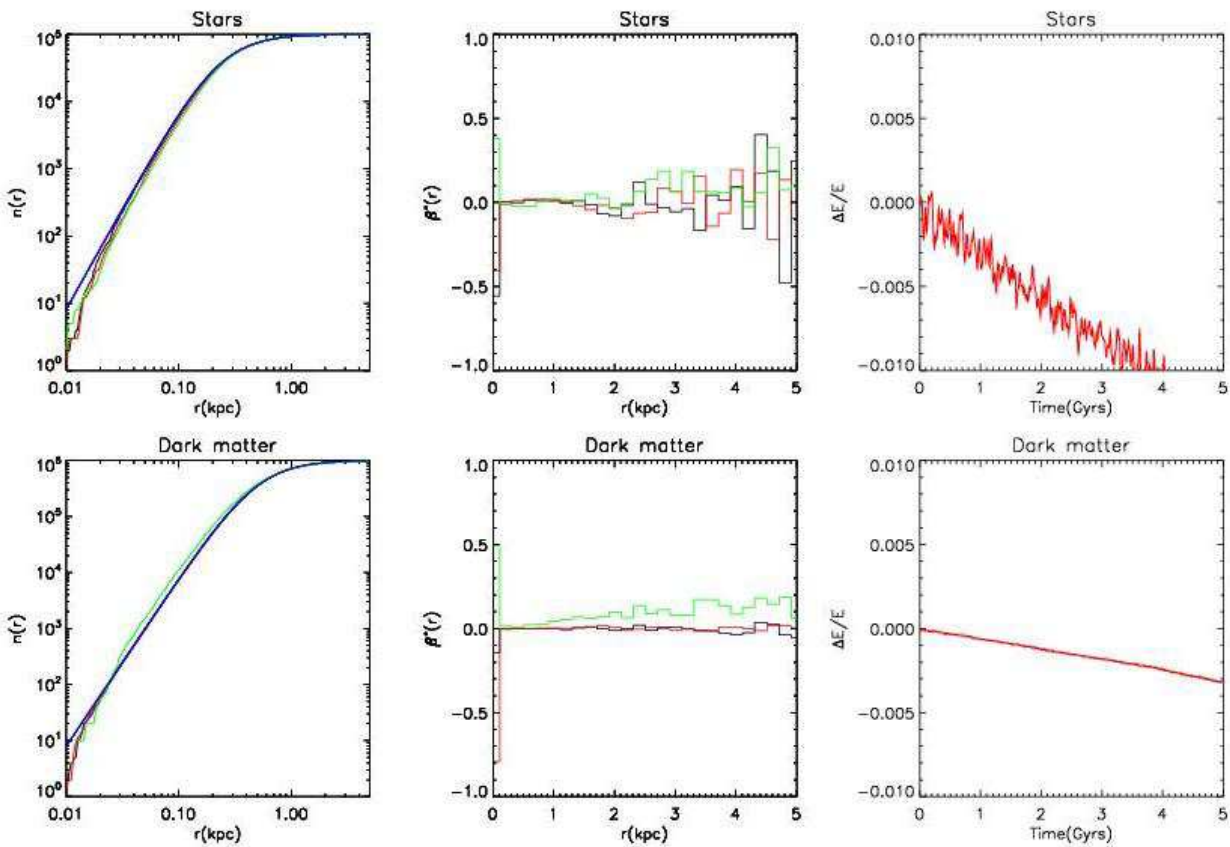,width=18cm}
\caption[]{Equilibrium tests for model A: The upper panels are for the
  stars, while the lower panels are for the
  dark matter. The left panels show the cumulative mass profile, 
  the middle panels show the velocity anisotropy and the right panels
  show the fractional change in energy as a function of time. The blue
  lines are for the analytic initial conditions, the black lines are for the
  simulated initial conditions with $10^6$ dark matter and $10^5$
  star particles, the red lines are for the simulated initial
  conditions evolved for 5\,Gyrs, while the green lines are for a
  simulation set up using the Maxwellian approximation
  \citep{1993ApJS...86..389H} and evolved for 5\,Gyrs.}
\label{fig:ictestmodela} 
\end{center} 
\end{figure*}

\subsection{Tidal shocking}\label{sec:shocking}

A second important effect of tides is tidal shocking. Tidal shocks can
be caused either as a satellite plunges through the disc of the host
galaxy (see e.g. \bcite{1972ApJ...176L..51O}) or moves on a highly
eccentric orbit through the galactic centre (see
e.g. \bcite{1987degc.book.....S}). Both scenarios are discussed in
detail by \citet{1997ApJ...474..223G}, \citet{1999ApJ...522..935G} and
\citet{1999ApJ...514..109G}.

In the impulsive limit, the mean energy injected into the satellite at
the r.m.s. radius, $\overline{r}$, for disc shocks is given by
\citep{1999ApJ...522..935G}:

\begin{equation}
\overline{\Delta E_\mathrm{disc}}=\frac{2 g_m^2 \overline{r}^2}{3V_z^2}(1+x^2)^{-\alpha}
\label{eqn:dEdisc}
\end{equation}
\noindent
where: $g_m$ is the maximal disc force along the $z$
direction (in all of the calculations presented in this paper
  it is a assumed that the plane of the Milky Way disc lies
  perpendicular to the 
  $z$-axis); $V_z$ is the $z$ component of the satellite velocity at
the point it passes through the disc; $x = \omega \tau$ ($\omega$ is
the angular velocity of a star within the satellite at $r$; $\tau \sim
2H/V_z$ is the typical shock time for a disc of scale height, $H$); and
$\alpha$ is the adiabatic correction exponent which is $\sim 5/2$ for
fast shocks \citep{1999ApJ...522..935G}.

Similarly, the mean energy injected for an eccentric orbit in a
spherical system (here an isothermal potential for the host
galaxy is assumed) is given by \citep{1999ApJ...514..109G}:

\begin{equation}
\overline{\Delta E_\mathrm{sph}}=\frac{1}{6}\left(\frac{v_0^2\pi
    \overline{r}}{R_{p}V_{p}}\right)^2(1+x^2)^{-\alpha}
\label{eqn:dEsph}
\end{equation}
\noindent
where $v_0$ is the asymptotic circular speed of test particles at
large radii in the host galaxy potential, $R_{p}$ is the
pericentric radius, $V_p$ is the satellite velocity at $R_p$ and
$x$ is as above, but the shock time is now given by, $\tau \sim \pi
R_p/V_p$.

We make the somewhat crude assumption that the total energy injected
into the satellite due to a single shock at $\overline{r}$ is then
given by the sum of the above two terms.

Assuming that the satellite is in virial equilibrium before the tidal
shock and that it has an isotropic velocity distribution, the total
specific energy of the satellite, $E$, is given by:

\begin{equation}
E = -T \sim \frac{1}{2}\overline{v^2(\overline{r})} =
\frac{1}{2\rho_s(\overline{r})}\int_{\overline{r}}^\infty {\rho_s(r) \frac{G M_s(r)}{r^2}}dr
\label{eqn:virialinit}
\end{equation}
\noindent
where $T$ is the satellite's total specific kinetic energy,
$\overline{v^2(\overline{r})}^{1/2}$ is the velocity dispersion, $\rho_s(r)$
and $M_s(r)$ are the mass and density profile of the satellite and the
right hand side follows from the Jeans equations
\citep{1987gady.book.....B}.

We assume that tidal shocks are important for the satellite if the
energy injected is comparable to the total energy of the satellite
initially. This gives: 

\begin{equation}
\gamma = 2\left[\frac{\overline{\Delta E_\mathrm{sph}}+\overline{\Delta E_\mathrm{disc}}}{\overline{v^2(\overline{r})}}\right]
\label{eqn:gammafac}
\end{equation}
\noindent
where $\gamma$ parameterises the importance of tidal shocks: tidal
shocks are important for $\gamma \sim 1$ and unimportant for $\gamma
\ll 1$. Notice from equations 
\ref{eqn:dEdisc}, \ref{eqn:dEsph} and \ref{eqn:gammafac} that the
effectiveness of tidal shocks depends on the mass density of the
satellite, the potential of the host galaxy and on the orbit of the
satellite. The strength of such shocks is very sensitive to the value
of $R_p$. For the Milky Way potential employed in this paper, both
disc shocks and spherical tidal shocks are negligible for orbits which
never come closer than $\sim 20$\,kpc.

The value of $\gamma$ for each of the
four models presented in this paper (labelled A-D) are given in Table
\ref{tab:initial}. Model D is a case of particular interest. For this
model, $\gamma \sim 1$ and so tidal shocks are very important. Yet
the tidal radius (as obtained using equation \ref{eqn:rtperi}) lies at
the edge of the light distribution. Thus it is possible 
that some of the Local Group dSphs could be in a regime where their
visible light is strongly affected by tidal shocks but only very weakly
perturbed by tidal stripping. 

\section{The numerical technique}\label{sec:numerical}

\subsection{Setting up the initial conditions}\label{sec:initial}

\subsubsection{The dSph galaxy}

The initial conditions for the dSph galaxy were set up in a similar fashion to
\citet{2004ApJ...601...37K}, but for two-component spherical galaxies
comprising a dark matter halo and some stars. Since this procedure is
non-trivial we briefly outline the major steps involved here:\\
\\
(i) The particle positions for each component are realised using
standard accept/reject techniques with numerically calculated
comparison functions and analytic density profiles
\citep{1992nrca.book.....P}.\\
\\
(ii) We use isotropic, spherically symmetric, distribution functions
\citep{1987gady.book.....B}. For each separate component, $i$, with 
density, $\rho_i$, in the total {\it relative}\footnote{The relative
  gravitation potential, $\psi = -\Phi+\Phi_0$, used here is as defined in
  \citet{1987gady.book.....B}.} gravitational potential of all
components, $\psi$, the distribution function is given by the
Eddington formula:

\begin{equation}
f(\epsilon) = \frac{1}{\sqrt{8}\pi^2}\left[\int_0^\epsilon
  {\frac{d^2\rho_i}{d\psi^2}\frac{d\psi}{\sqrt{\epsilon-\psi}}+\frac{1}{\sqrt{\epsilon}}\left(\frac{d\rho_i}{d\psi}\right)_{\psi=0}}\right]
\label{eqn:eddington}
\end{equation}
\noindent
where $\epsilon$ is the specific energy. Note that, for any choice of
$\rho_i$ and $\psi$ which are well behaved as $r \rightarrow
\infty$, the second term in the brackets in equation \ref{eqn:eddington}
vanishes \citep{1987gady.book.....B}.

\begin{figure*} 
\begin{center}
\epsfig{file=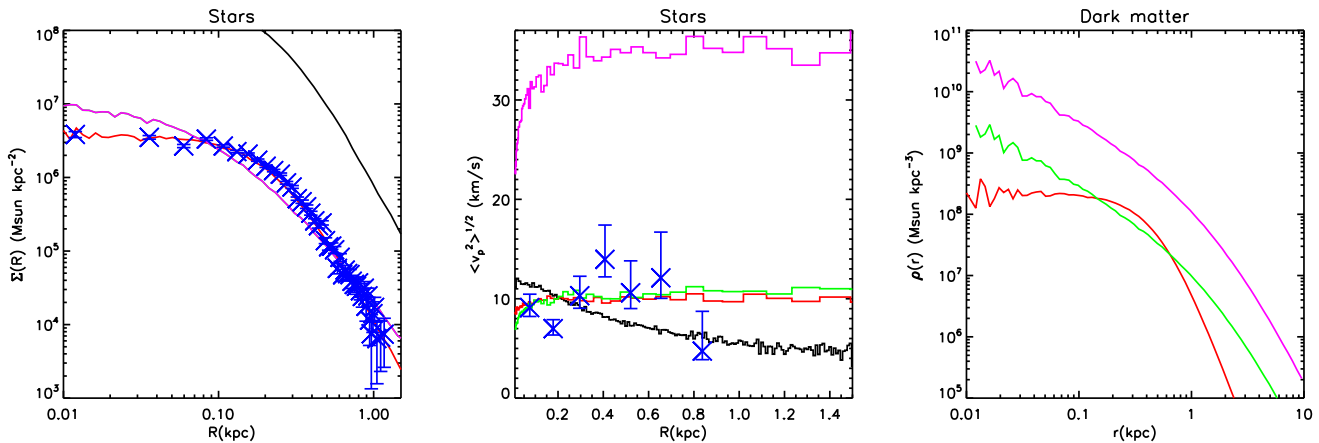,width=18cm}
\caption[]{Initial conditions for models A-D. The blue data points
  are for the Draco dSph galaxy taken from
  \citet{2004ApJ...611L..21W}. The black,red,green,purple lines are
  for models A,B,C,D respectively. Recall that in model A we have no
  dark matter, hence there is no black line in the right panel plot of
  the dark matter distribution.}
\label{fig:initial}
\end{center}
\end{figure*}

For a general two component system $f(\epsilon)$ must be calculated
numerically. In \citet{2004ApJ...601...37K}, where only single
component systems (dark matter only) were considered, the term
$\frac{d^2\rho_i}{d\psi^2}$ could be calculated
analytically\footnote{In fact, even for single component models,
  $\frac{d^2\rho_i}{d\psi^2}$ is only analytic for a few special
  cases.}. Here,
where in general $\psi = \psi_1+\psi_2+...+\psi_n$ for an $n$ component
system, this must be calculated numerically. We consider the special
case where $\psi_i$ and $\rho_i$ are known analytically for each
component. In this case we can write:

\begin{equation}
\frac{d^2\rho_i}{d\psi^2} =
\left(\frac{dr}{d\psi}\right)^2\frac{d^2\rho_i}{dr^2}
+ \frac{dr}{d\psi}\frac{d\rho_i}{dr}\frac{d}{dr}\left(\frac{dr}{d\psi}\right)
\end{equation}
\noindent

This means that we can calculate $\frac{d^2\rho_i}{d\psi^2}$ in
general {\it analytically} for a given value of $r$. This avoids the
need for a noisy double numerical differential. However, we do
still need to generate a look-up table to solve the inversion for
$r(\psi)$. Once this is done, however, we may perform the integral in
equation \ref{eqn:eddington} and create a log look-up table for $f(\epsilon)$
over a wide range of $\epsilon$ as in \citet{2004ApJ...601...37K}.\\
\\
(iii) Once the distribution function, $f(\epsilon)$, is known for each
component, we can set up particle velocities using standard
accept/reject techniques \citep{1992nrca.book.....P}. We take
advantage of the fact that the maximum value of the
distribution function at a given point will be given by the maximum
value of $\epsilon$; that is for $v=0$ \citep{1995MNRAS.277.1341K}.\\
\\
This may seem like an unnecessarily large amount of effort to go to,
when simple numerical techniques already exist for setting up multi-component
galaxies \citep{1993ApJS...86..389H}. However, as pointed out by
\citet{2004ApJ...601...37K}, using approximate methods such as the
Maxwellian approximation will lead to a small evolution away from the
initial conditions. This can lead to significant radial velocity
anisotropies being introduced at large radii. In a companion paper
\citep{Readinprep1} we show that such anisotropies lead to
significantly increased tidal stripping. This is undesirable
when we wish to perform controlled numerical simulations of tidal stripping.

In Figure \ref{fig:ictestmodela} we show an equilibrium test for
our two-component model B (the other models showed similar results). The upper
panels are for the stars, while the lower
panels are for the dark matter. The left panels show the cumulative
mass profile, the middle panels show the velocity anisotropy and the
right panels show the fractional change in energy as a function of
time. We use a modified anisotropy parameter, $\beta_*$, given by:

\begin{equation}
\beta_* = \frac{\overline{v_r^2}-\overline{v_t^2}}{\overline{v_r^2}+\overline{v_t^2}}
\label{eqn:betastar}
\end{equation}
\noindent
where $\overline{v_r^2}$ and $\overline{v_t^2}$ are the radial and
tangential velocity dispersions respectively. This definition gives
$\beta_*=-1$ for pure circular orbits, $\beta_*=0$ for isotropic
orbits, and $\beta_*=1$ for pure radial orbits\footnote{The standard
  anisotropy parameter, $\beta$, is usually defined as $\beta = 1 -
  \overline{v_\theta^2}/\overline{v_r^2}$
  \citep{1987gady.book.....B}. This definition is useful for obtaining
  analytic solutions to the Jeans equations. However, it can be
  readily seen that for circular orbits $\beta \rightarrow -\infty$. In
  this paper it is useful to have a definition of anisotropy which is
  well-behaved for all orbits and which is symmetrical around
  isotropic orbits.}.

The blue lines in Figure \ref{fig:ictestmodela} are for the analytic
initial conditions, the black lines are for the simulated initial
conditions with $10^6$ dark matter and $10^5$ star particles,
the red lines are for the simulated initial conditions evolved in
isolation (i.e. with no external tidal field) for
5\,Gyrs, while the green lines are for a simulation set up using the
Maxwellian approximation \citep{1993ApJS...86..389H} and evolved for
5\,Gyrs.

The most important point to take away from Figure
\ref{fig:ictestmodela} is that we can confirm the findings of
\citet{2004ApJ...601...37K} for our two-component model. In the
Maxwellian approximation the dark matter evolves away rapidly and
significantly from the initial conditions. In particular, significant
radial anisotropy is introduced to the dark matter at large radii (see
green line, bottom right panel). In contrast, the distribution
function generated initial conditions are very stable over the whole
simulation time.

Notice that the energy both in the stars and dark matter is conserved
to better than 2\% over the whole simulation time.

\subsubsection{The host galaxy potential}

We used a host galaxy potential chosen to provide a good fit to the Milky Way
\citep{2005ApJ...619..807L}, with a
Miyamoto-Nagai potential for the Milky Way disc and 
bulge \citep{1976PASJ...28....1N}, and a logarithmic potential for the
Milky Way dark matter halo. These are given by respectively
\citep{1987gady.book.....B}:

\begin{equation}
\Phi_\mathrm{mn}(R,z) = \frac{-G M_d}{\sqrt{R^2+(a+\sqrt{z^2+b^2})^2}}
\label{eqn:miyamoto}
\end{equation}
\noindent
where $M = 5 \times 10^{10}$M$_\odot$ is the disc mass, $a=4$\,kpc is
the disc scale length and $b=0.5$\,kpc is the disc scale height, and:

\begin{equation}
\Phi_{\mathrm{log}}(r) =
\frac{1}{2}v_0^2\ln\left(R_c^2+r^2\right)+\mathrm{constant} 
\label{eqn:loghalo}
\end{equation}
\noindent
where $R_c=4.1$\,kpc is the halo scale length 
and $v_0=220$km/s is the asymptotic value of the circular speed of
test particles at large radii in the halo.

Note that with the choice $b=0.5$\,kpc, we do not consider the case of
maximal disk shocking which would occur for $b \rightarrow 0$ (see
equation \ref{eqn:dEdisc}). However, given that the amount of disc
shocking is also degenerate with the orbit of the dSph which is poorly
constrained, our simulations span a range from typical to extreme
levels of shocking. In addition, none of our conclusions would be
altered by stronger disk shocks.

\subsection{Force integration and analysis}

The initial conditions were evolved using a version of the
GADGET N-body code \citep{2001NewA....6...79S} modified to include a
fixed potential to model the host galaxy.

A wide range of code tests were performed using different softening
criteria, force resolution and timestep criteria and the results were
found to be in excellent agreement with each other. We also explicitly
checked the code against 
output from two other N-body codes: \citet{2000ApJ...536L..39D} and
 NBODY6 \citep{1999PASP..111.1333A}, and found excellent agreement. Our force
softening was chosen using the criteria of
\citet{2003MNRAS.338...14P} and is shown in Table
\ref{tab:initial}. For the equilibrium tests, our simulations were
found to conserve {\it total} energy to better than one part in $10^3$ over the
whole simulation time of 5\,Gyrs.

Finally, it is not trivial to mass and momentum centre the satellite
when performing analysis of the numerical data. An incorrect mass
centre can lead to spurious density and velocity features
\citep{Readinprep3}. We use the method of shrinking spheres to
find the mass and momentum centre of the satellite
\citep{2003MNRAS.338...14P}.

\begin{figure*} 
\begin{center}
\epsfig{file=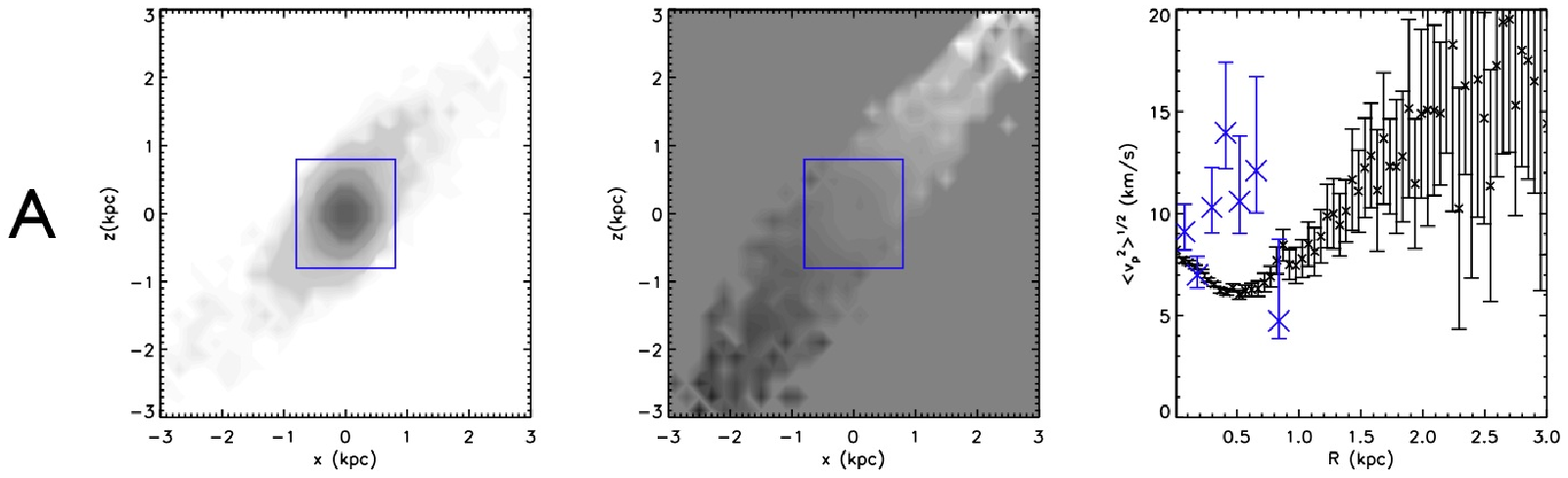,width=15cm}\\
\epsfig{file=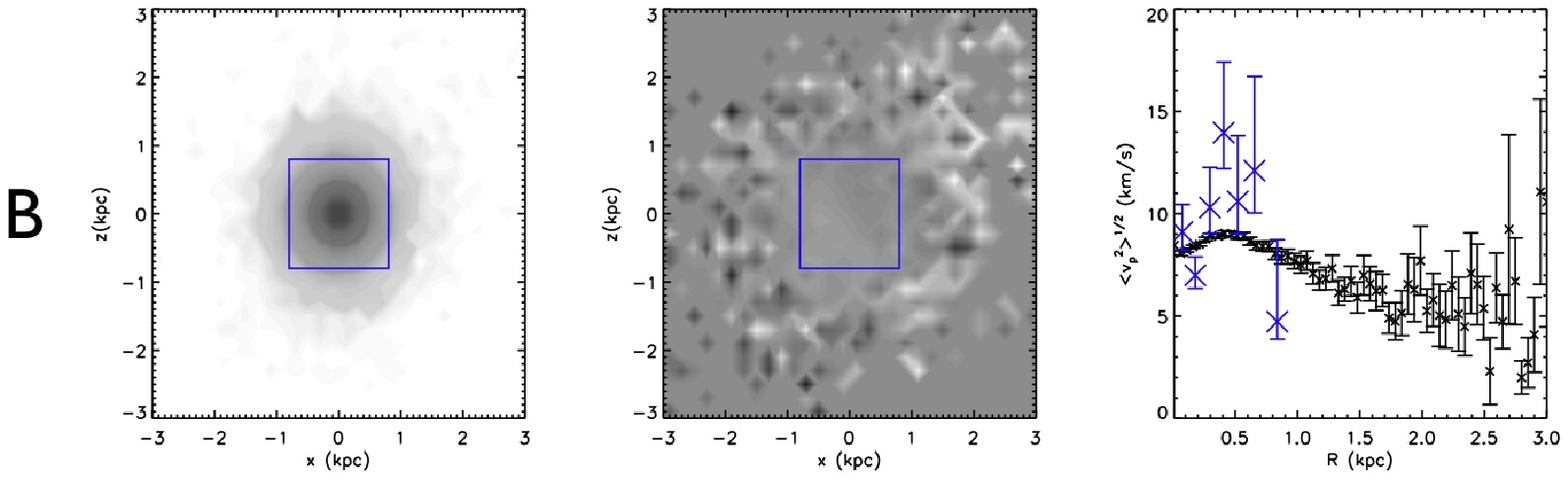,width=15cm}\\
\epsfig{file=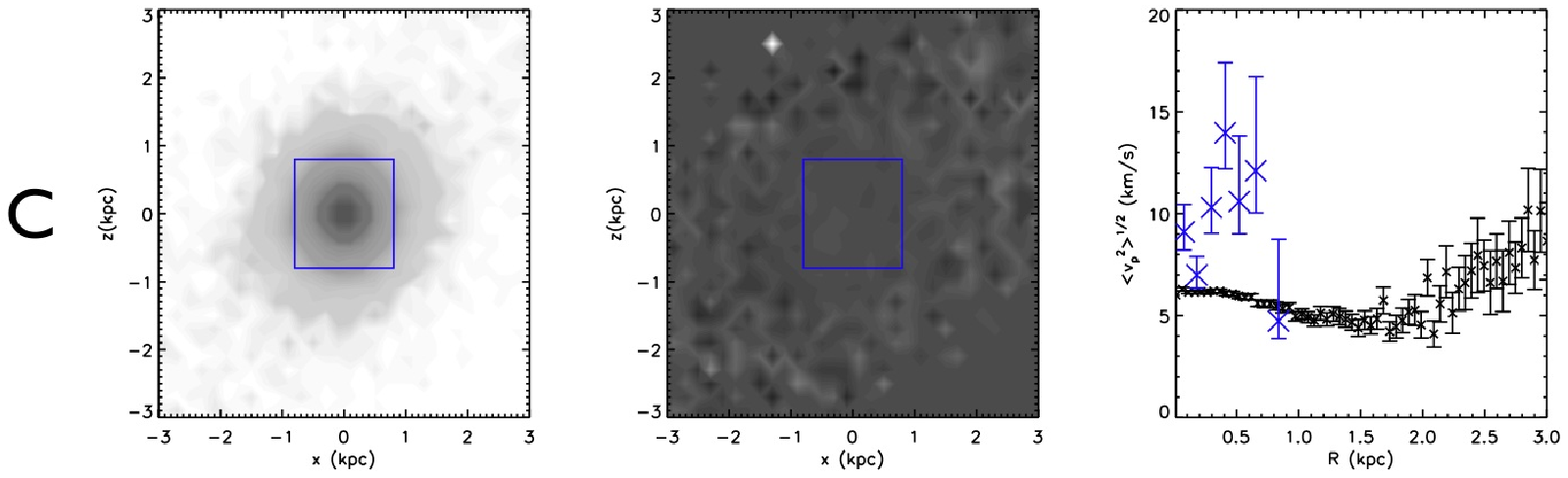,width=15cm}\\
\epsfig{file=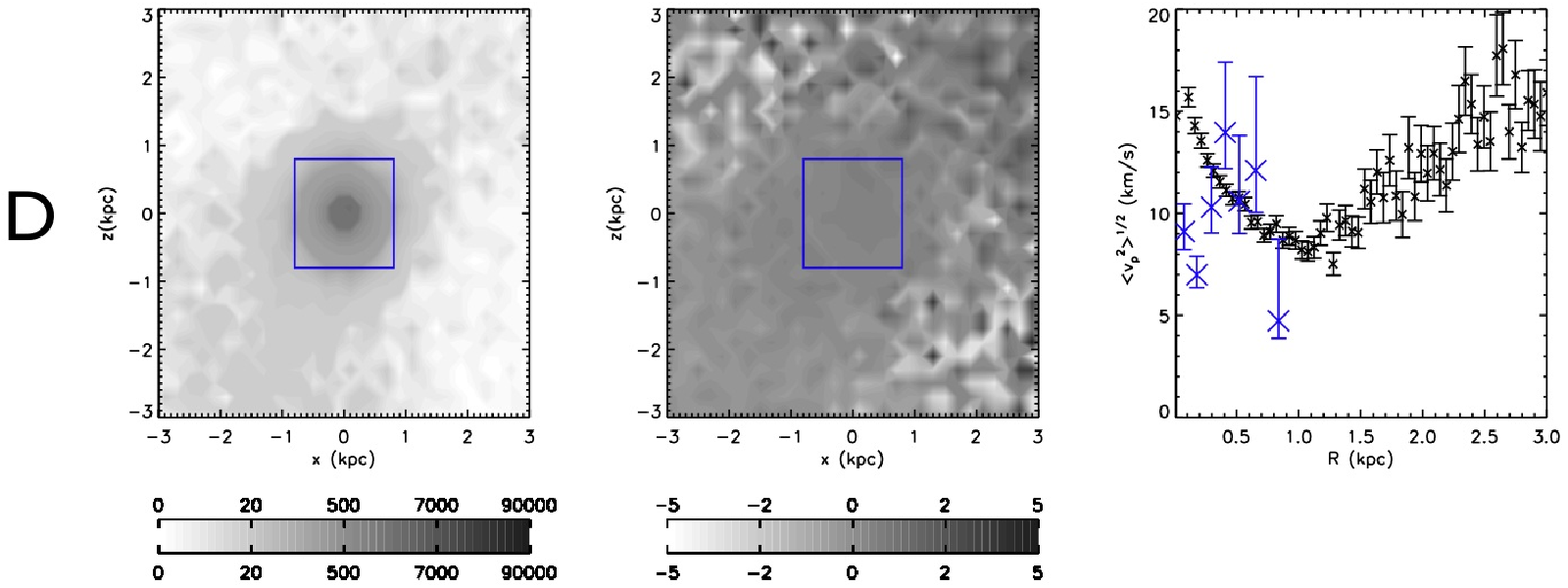,width=15cm}
\caption[]{Projected surface density (left), velocity (middle) and
  velocity dispersion (right) for models A (top) through D
  (bottom). The grey contours show particle numbers per unit area (left) and
  velocities in km/s (middle). Over-plotted on the projected velocity
  dispersion plot (right) are the data for the Draco dSph galaxy taken
  from \citet{2004ApJ...611L..21W}. The blue boxes overlaid on the
  contours mark approximate current observational limits. The
  projections shown are typical; different projections for model A are
  shown in Figure \ref{fig:proj}.}
\label{fig:results}
\end{center}
\end{figure*}

\subsection{The choice of initial conditions: models A-D}\label{sec:initialAD}

We chose four models for our initial conditions labelled A-D as shown
in Table \ref{tab:initial} and Figure \ref{fig:initial}. In all models
the density distribution for the stars and dark matter were given
either by Plummer spheres \citep{1987gady.book.....B} or by Split
Power law profiles (SP) (\bcite{1990ApJ...356..359H},
\bcite{1992MNRAS.254..132S}, \bcite{1993MNRAS.265..250D} and
\bcite{1996MNRAS.278..488Z}). The 
density-potential pairs for these distributions are given by respectively: 

\begin{equation}
\rho_\mathrm{plum} = \frac{3M}{4\pi a^3}\frac{1}{(1+\frac{r^2}{a^2})^{5/2}}
\label{eqn:rhoplum}
\end{equation}
\begin{equation}
\Phi_\mathrm{plum} = \frac{-G M}{\sqrt{r^2+a^2}}
\label{eqn:psiplum}
\end{equation}

\begin{equation}
\rho_\mathrm{SP} = \frac{M(3-\alpha)}{4\pi a^3}\frac{1}{(r/a)^\alpha(1+r/a)^{4-\alpha}}
\label{eqn:rhosp}
\end{equation}
\begin{equation}
\Phi_\mathrm{SP} = \frac{G M}{a(2-\alpha)}\left[(1+a/r)^{\alpha-2}-1\right]
\label{eqn:psisp}
\end{equation}
\noindent
where $M$ and $a$ are the mass and scale length in both cases, and
$\alpha$ is the central log-slope for the SP profile (note that the SP
profile always goes as $\rho_\mathrm{SP} \propto r^{-4}$ for $r \gg
a$).

As detailed in section \ref{sec:introduction}, each of these models
was chosen to test a different scenario for the evolution of dSphs in
the presence of tides. In all of the models, the dSph was placed on a
plunging orbit which took it through the Milky Way disc. This ensured
that both tidal stripping and shocking were important (see
section \ref{sec:theory}). The initial conditions, orbit and output
time for each of the models are given in Table
\ref{tab:initial}. Models B and C placed the dSph on an orbit which is
consistent with current constraints from the proper motion of Draco
\citep{2001ApJ...563L.115K}. However, these proper motion measurements
are notoriously difficult and the errors large (see
e.g. \bcite{2002AJ....124.3198P}). Thus in models A and
D, we also consider more extreme (i.e. more radial) orbits.

\begin{figure*} 
\begin{center}
\epsfig{file=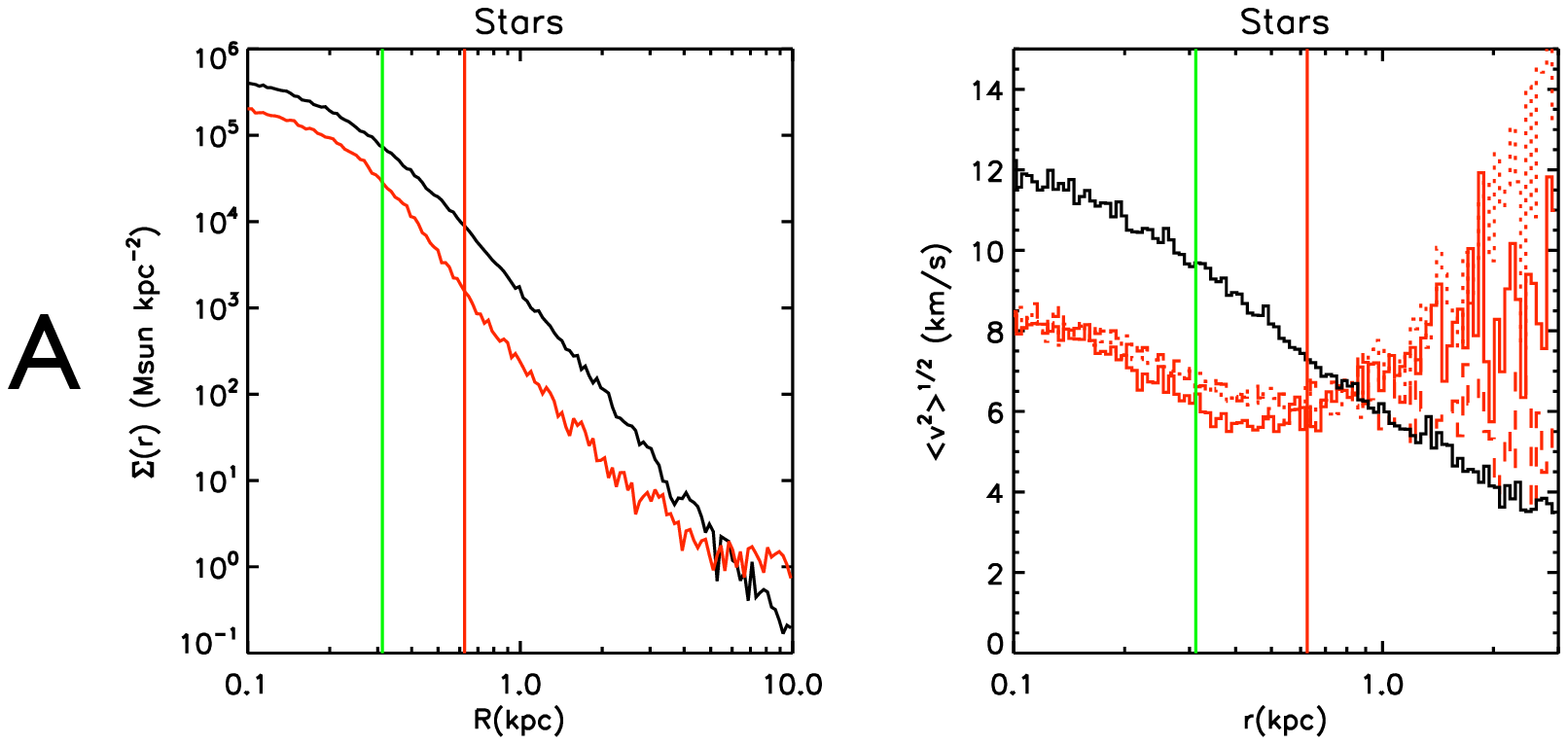,height=15cm,angle=-90}\\
\epsfig{file=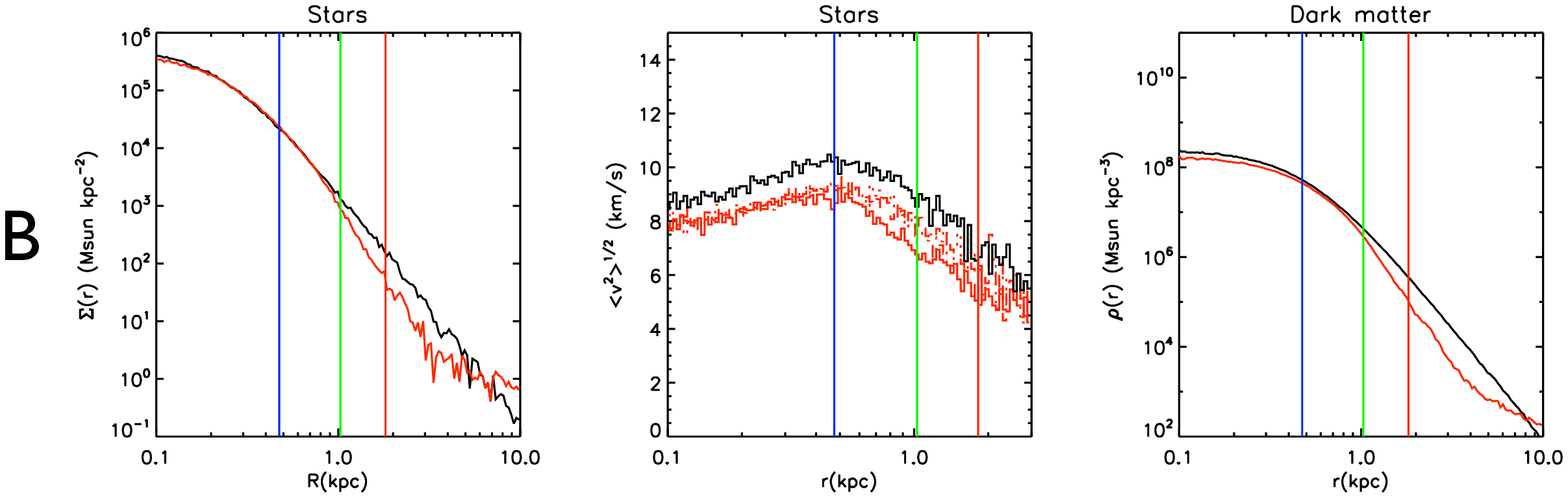,height=15cm,angle=-90}\\
\epsfig{file=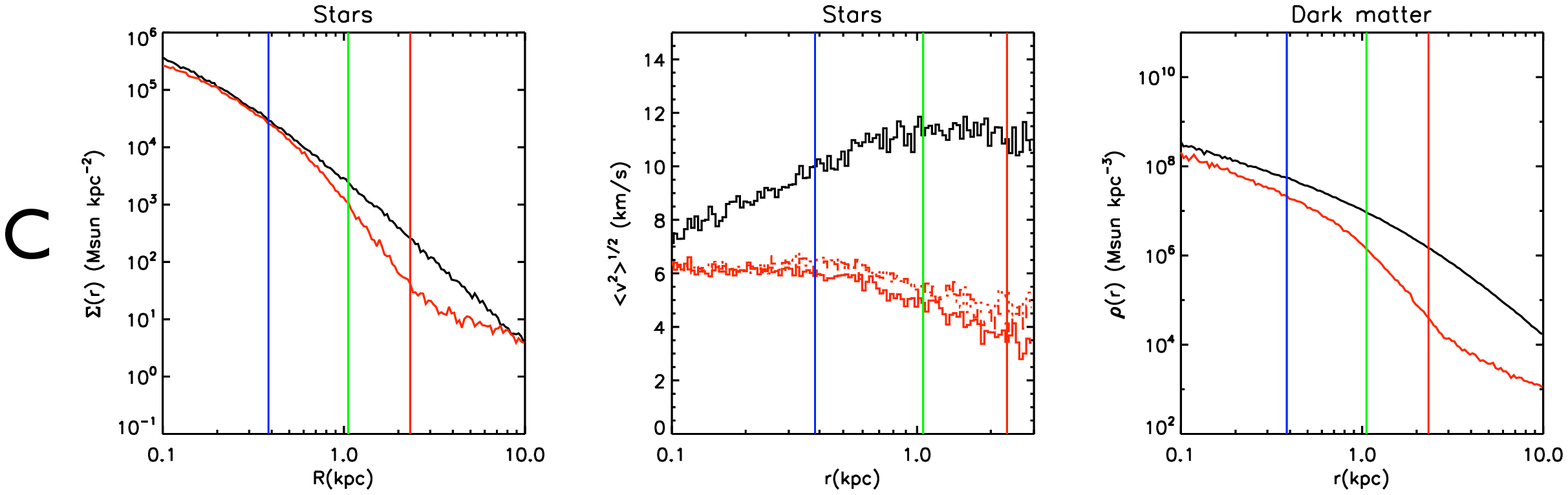,height=15cm,angle=-90}\\
\epsfig{file=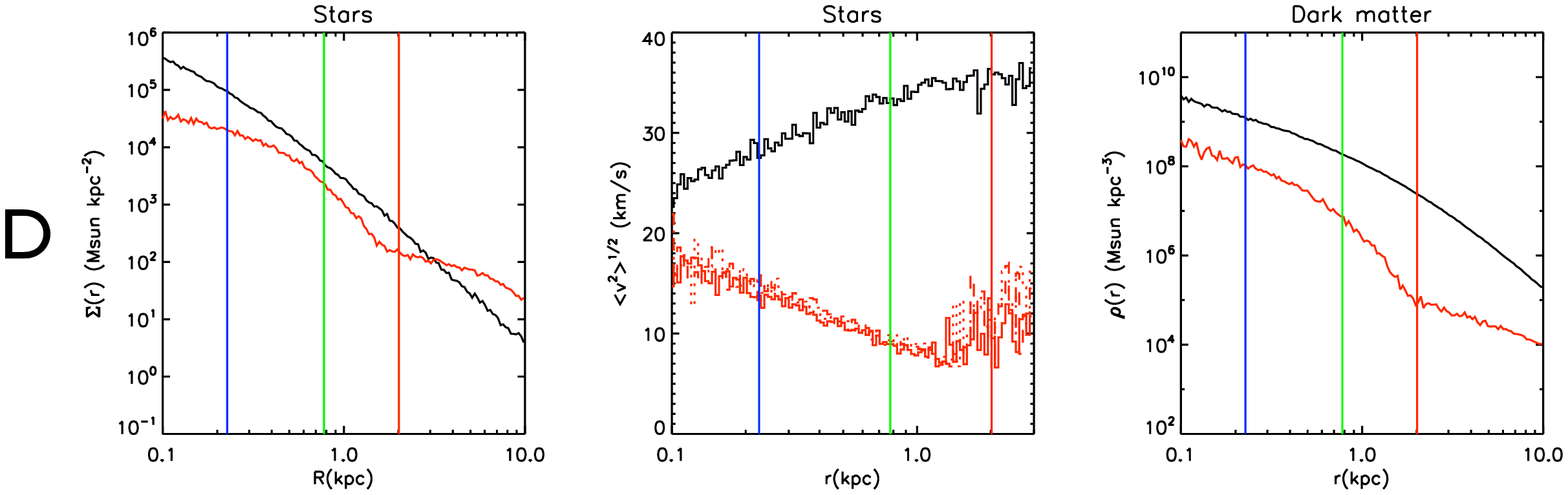,height=15cm,angle=-90}
\caption[]{Projected surface brightness profile
of the stars (left panel), velocity dispersion of the stars
(middle panel) and the density profile of the dark matter (right
panel). Over-plotted are three theoretical tidal radii calculated as
described in \citet{Readinprep1}; the blue/green/red vertical lines are for the
prograde/radial/retrograde tidal radii respectively. The black lines
show the initial conditions, while the red lines show the evolved
profiles at the output times given in Table \ref{tab:initial}. In the velocity
dispersion plot, the solid, dotted and dashed lines show the $r$,
$\theta$ and $\phi$ components of the velocity dispersion
respectively. Models A-D are shown from top to bottom. Model A did not
contain any dark matter which is why no plot is shown for the dark
matter in this case. The tidal radius for prograde star orbits
(vertical blue line) cannot be seen for model A since it lies at very
small radii (9\,pc).}
\label{fig:resultsden}
\end{center}
\end{figure*}

\begin{figure*} 
\begin{center}
\epsfig{file=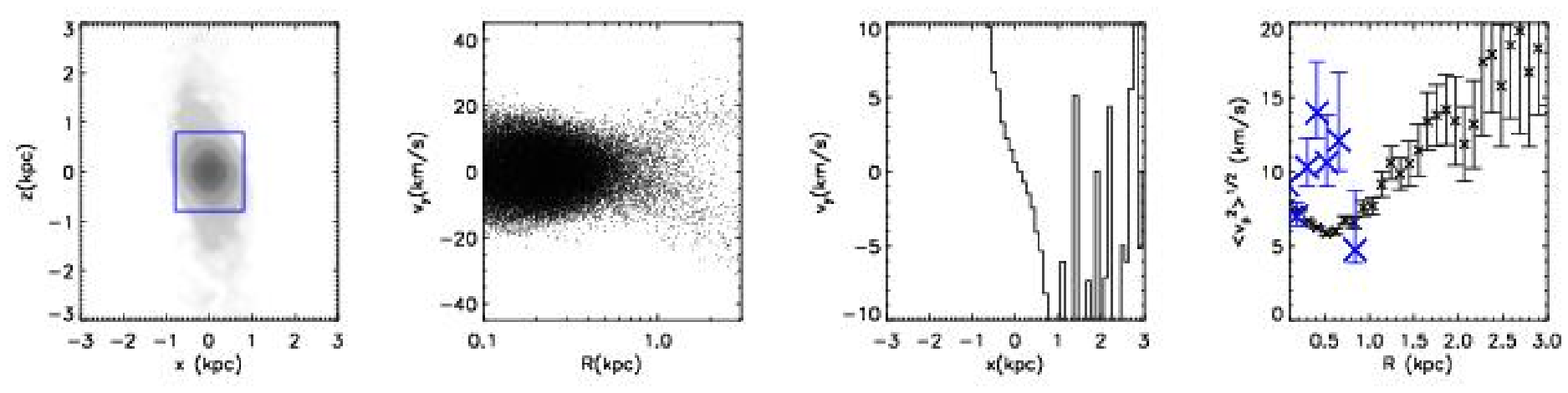,width=18cm}\\
\epsfig{file=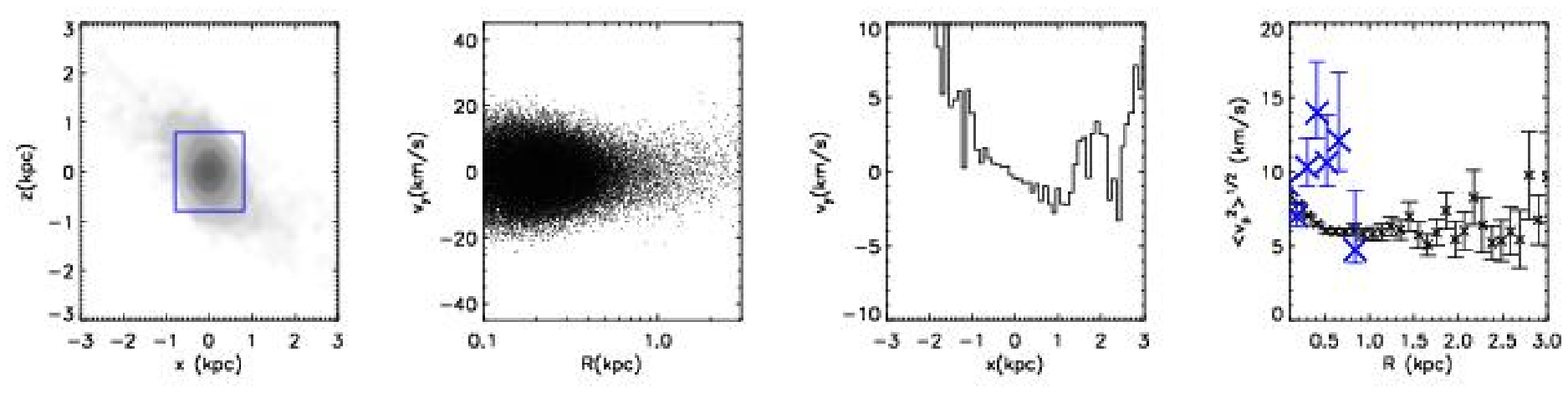,width=18cm}
\caption[]{The effect of viewing projection for model A. Top panels
  show a typical projection; bottom panels show a rare projection
  which kinematically hides one of the tidal tails. From left to right
  panels show: 
  projected surface density, individual projected star velocities, mean
  projected velocity and projected velocity dispersion. The grey
  contours show particle numbers 
  per unit area as in Figure \ref{fig:results}. Over-plotted on the
  projected velocity dispersion plot (right) are the data for the
  Draco dSph galaxy taken from \citet{2004ApJ...611L..21W}. The blue
  boxes overlaid on the contours mark approximate current
  observational limits. Note that the velocity gradients shown here
  are averaged over the $y$ and $z$ coordinates and are necessarily
  different to the two dimensional gradients shown in Figure
  \ref{fig:results}.}
\label{fig:proj}
\end{center} 
\end{figure*}

\section{Results}\label{sec:results}

The results for models A-D are presented in Figures \ref{fig:results}
and \ref{fig:resultsden}. In Figure \ref{fig:results}, we show the
projected surface brightness (left panel), projected velocity (middle
panel) and projected velocity dispersion\footnote{Note that throughout
  this paper, all velocity dispersions are calculated using the {\it
    local} mean velocity; this is in contrast to observational techniques which
  often use the global mean for all the stars. The use of the global
  mean is only justified where there are no velocity gradients
  measured; this is the case for the observations. It is not the case,
  however, for some of the models presented here where velocity
  gradients are induced through tides.} (right panel) for the
stars. For the projected 
velocity dispersion, the errors are determined from Poisson
statistics. Models A-D are presented down the page. In Figure
\ref{fig:resultsden} we show the projected surface brightness profile
for the stars (left panel), the true velocity dispersion of the stars
(middle panel) and the density profile of the dark matter (right
panel). For clarity on the plots, we compare the results of our
simulations to data from the Draco dSph only, on the assumption that Draco is
representative of the most dark matter dominated Local Group dSphs
(see section \ref{sec:introduction} and Figure
\ref{fig:localdata}). In Figure \ref{fig:proj} we show the
effect of different viewing angles in projection on the sky, using
model A as an example. In Figure \ref{fig:massloss} we show the mass
loss, due to tides, as a function of time for all four models A-D.

\subsection{The no-dark matter hypothesis: model A}\label{sec:modela}

In model A, we test the hypothesis that we can eliminate the need for
dark matter in dSph galaxies by instead employing tidal heating to
raise the velocity dispersions. One
of the appealing aspects of such a model is that it produces tidally
distorted isophotes, such as those which may have been seen in some dSphs
(see e.g. \bcite{2001ApJ...549L..63M} and
\bcite{2003AJ....125.1352P}).

In Figure \ref{fig:results}, top left panel, we can see that 
distortions occur naturally in such a tidal model and the S-shaped
isophotes caused by tides can be clearly seen in this
projection. However, once the kinematic data are taken into account, it
is clear that such a model runs into serious difficulties. The
projected velocity shows a strong gradient (10\,km/s/kpc) due to the
near-dominant 
tidal arms, whereas no such velocity gradients have been observed in
the Local Group dSphs so far \citep{2004ApJ...611L..21W}. Furthermore,
the projected velocity dispersion rises out towards the edge of the
light, whereas the Local Group dSphs like Draco have flat or falling
projected velocity dispersions (see e.g. \bcite{2004ApJ...611L..21W},
\bcite{2005ApJ...631L.137M}, \bcite{2004MNRAS.354L..66K} and Figure
\ref{fig:localdata}). It is possible, however, that such tidal
features could be hidden in projection. Figure \ref{fig:proj}
shows the effect of different viewing 
projections on the sky for model A. The top and bottom panels show two
different projections: the top is a more typical projection, while the
bottom, which kinematically hides one of the tidal tails, is more
rare. By rare, we mean that such obscurations only occur for $\sim
1/5$ of all projection angles. While the orbit of any given dSph
would rule out some projection angles, when studying the generic
effect of tides on all of the Local Group dSphs, it seems reasonable
to consider such random projections. 

Notice from the middle left panels, that the primary cause of
the rising velocity dispersion in projection is the cylindrical
averaging which picks up 
both tidal tails. This is why, when a suitable projection is found
which hides one of the tails, the velocity dispersion becomes flat
rather than rising (bottom panels). In this case the velocity
gradients are visible (middle right panel) only at large radii. Interior to
$\sim 1$\,kpc, the velocity gradient is less than $\sim 5$\,km/s. Such
a small velocity gradient would be difficult to detect
observationally \citep{2004ApJ...611L..21W}. However, even this
special projection is inconsistent with the data from Draco, UMi
and Sextans which all show {\it falling} velocity dispersions in
projection (see Figure \ref{fig:localdata}).

It is worth recalling at this point that we are not trying to explicitly
model the Draco, UMi or Sextans dSphs. Instead we show that the generic effect
of tides is to produce velocity gradients and flat or rising projected
velocity dispersions. That this is not seen in Draco, Sextans or UMi
suggests that while tidal stripping may be important beyond the edge
of the currently observed light ($\sim$1\,kpc), it is unlikely to have been
important within this radius.

An interesting feature of model A is that the small pericentre of the
satellite (9\,kpc) causes strong tidal shocks which over 10\,Gyr lower
the central density of the stars by a factor of $\sim 2$. This can be
seen in Figure \ref{fig:resultsden}, top panels. These tidal shocks,
as discussed in \citet{Readinprep1} wash out the tidal 
features which would otherwise be expected at the analytic tidal radii
(blue, red and green vertical lines in Figure
\ref{fig:resultsden}). However, the effects of tides can still be seen
in the velocity dispersions. Notice that tangential anisotropy, which
should be present beyond the prograde stripping radius is
present at all radii over the satellite (see section
\ref{sec:stripping}). For model A, the prograde
stripping radius lies at just 9\,pc and hence does not show up in
Figure \ref{fig:resultsden}.

\begin{figure*} 
\begin{center}
\epsfig{file=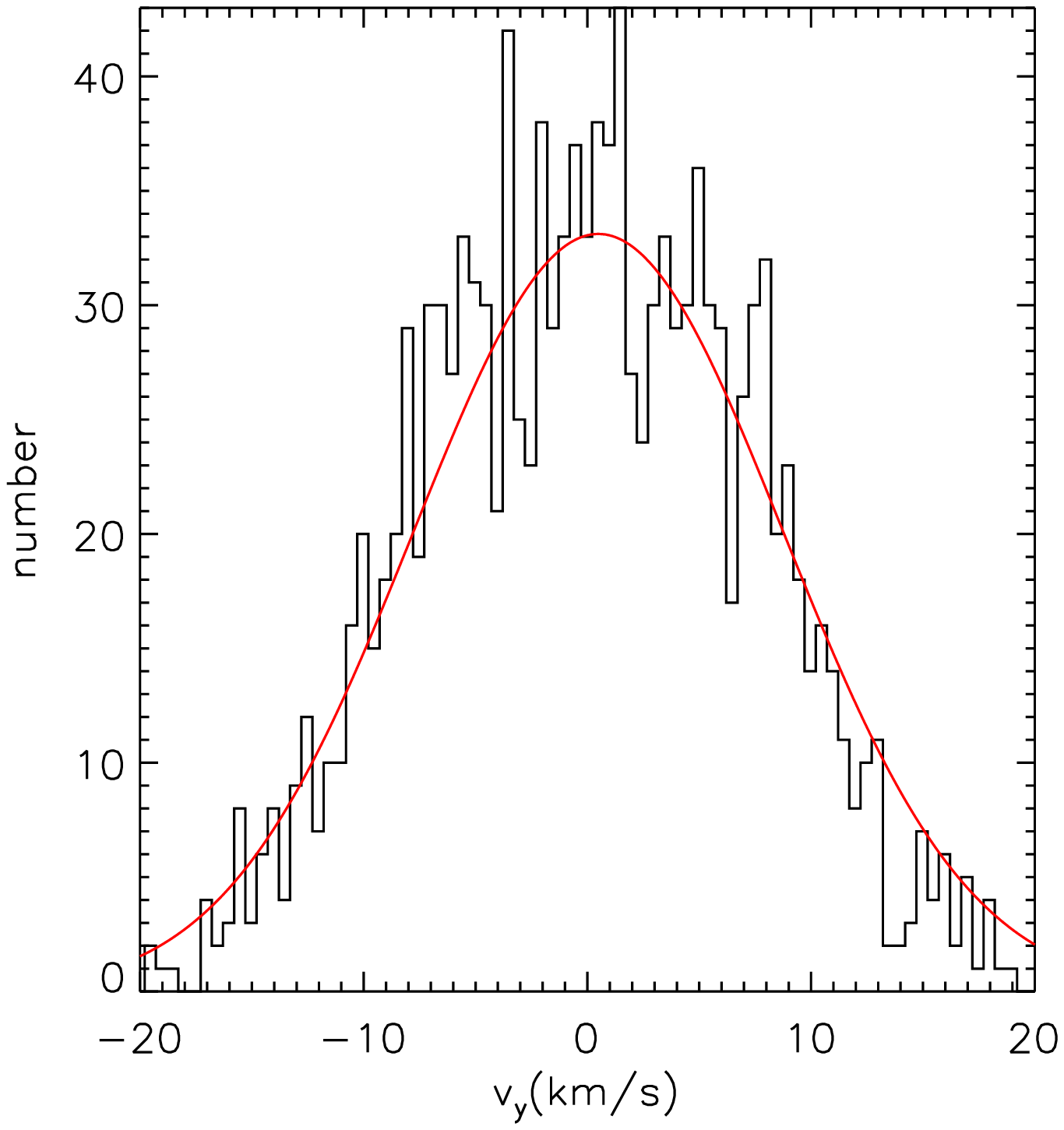,width=8.5cm}
\epsfig{file=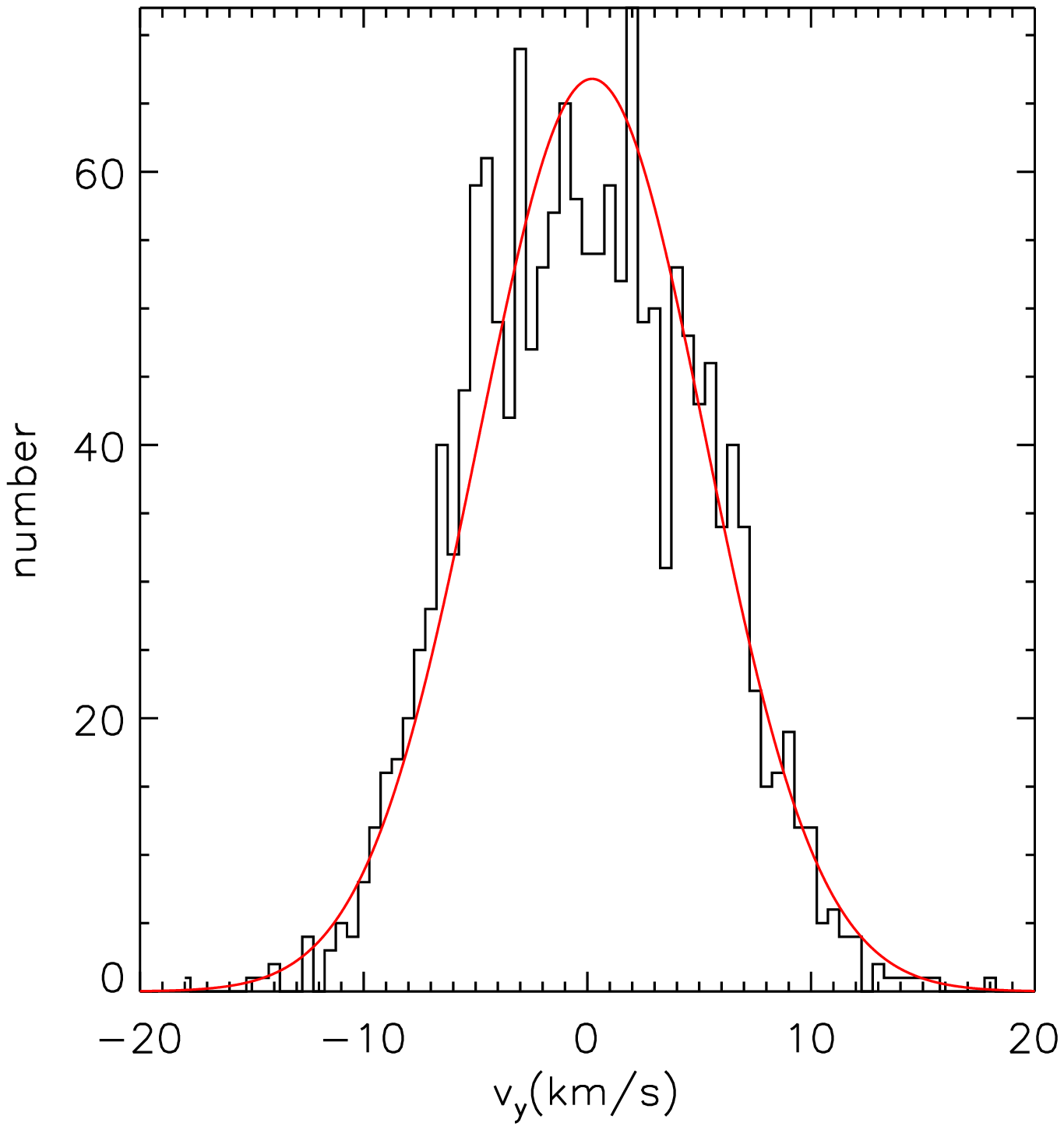,width=8.5cm}
\caption[]{Histogram of the stellar velocities in a 0.2\,kpc bin at
  1\,kpc for model B (left) and model C (right). Over-plotted are
  Gaussian fits in both cases (red lines).} 
\label{fig:histobin}
\end{center} 
\end{figure*}

\subsection{Weak tides: models B and C}\label{sec:modelbc}

In models B and C we consider the effect of weak tides on the Local
Group dSph galaxies. In this case we imagine that tides gently shape
the outer regions of dSphs over a Hubble time.

In Figure \ref{fig:results}, second and third panels from top, we can
see that tides have little affected models B and C. The surface
brightness distributions are nearly spherical, as in the initial
conditions, and there are no visible velocity gradients across the
galaxy. Tidal heating of the outermost stars is present in both models
from about $\sim 2$\,kpc outwards. Notice that in neither model is there
a drop in the projected velocity dispersion. The decline in model A
was present in the initial conditions and is not as sharp as that seen
in the data from Draco.  

This can also be seen in Figure \ref{fig:resultsden}, second and third
panels from top. Notice that, as in model A, tangential velocity
anisotropy appears at the prograde stripping radius (vertical blue
line) in both models B and C. Notice also the action of tidal shocks,
particularly on the dark matter in model C. As the satellite is puffed
up by shocks, its velocity dispersion at all radii is lowered. This
may seem counter-intuitive at first: the satellite is heated by tidal
shocks and yet its velocity dispersion falls. However, it is the
subsequent expansion of the satellite after the injection of
energy from the tidal shock which causes a drop in its velocity
dispersion {\it at a given radius}\footnote{In
  projection this becomes a little more complicated since hot stars
  which have moved out to large radii can inflate the projected
  velocity dispersion. This does not happen in practice, however,
  because these stars move beyond the tidal stripping radius,
  become unbound and move away from the satellite.}.

Notice further that it is only in projection that the velocity
dispersions appear hot beyond the tidal radius. The true velocity
dispersions are still gently falling, as in the initial conditions, at
this point. This occurs for two reasons. First, and this is usually
the dominant effect for a given projection, the cylindrically averaged
tidal tails cause the projected velocity dispersion to rise (see
Figure \ref{fig:proj}). Second,
the onset of tangential anisotropy makes the outermost stars appear hot in
projection. This second effect is only important when the tidal tails
are viewed along the line of sight, thereby minimising the extent to
which they can be observed in projection.

This is interesting: in all models tidal stripping leads to a rising
or flat projected velocity dispersion. There is never a sharp decline
such as that observed in the Draco, UMi and Sextans dSphs
(see section \ref{sec:introduction}). We now discuss why tidal stripping
always leads to hot stars in projection by considering three
mechanisms by which tidal stripping might have led instead to the
opposite - namely a cold outer point in the projected velocity
dispersion.\\
\\
{\it (i) The outer-most point is the result of a chance projection
  effect between the galaxy and its tidal tail along the line of
  sight.}\\
\\
We explicitly checked for this 
by searching over 5 degree projections every half orbital period for
both models B and C and found no sharp features at any output
time. If a projection 
effect could cause such a sharp drop it would be very rare and, while
one could envisage this for one Local Group dSph galaxy, the fact that
such a sharp feature has been observed in at least three renders this
possibility unlikely.\\
\\
{\it (ii) The dSph galaxy is on a significantly plunging orbit. The
  outermost stars are heated up by a disc shock as the dSph plunges
  through the Milky Way disc. The dSph galaxy then moves out towards
  apocentre and the hot outer stars escape leaving a cold population
  behind.}\\
\\
This hypothesis was presented in \citet{2004ApJ...611L..21W} and
formed one of the motivations for this current work. The output time
chosen for models B and C has been specifically selected to test this
hypothesis. We choose a time when the dwarf galaxy has 
spent a large amount of time away from the Milky Way disc and the
outermost stars heated by the last passage have had time to
escape. The hypothesis is falsified for three reasons. First, for the
case of models B and C, tidal shocking replenishes stars beyond the
tidal radius and so there is always a source of hot tidal stars which
mask any cold populations. Secondly, the image of a sharp edge to the
satellite beyond which a cold population of stars might exist is not
correct. As shown in a companion paper (\bcite{Readinprep1}; and see
also section \ref{sec:theory}), the tidal radius of the satellite
depends on the orbit of the stars within the satellite. This leads to a
continuum of tidal radii and, therefore, no sharp edge to the
satellite. Finally, even if there were such a sharp edge, and
tidal shocking did not refill stars beyond the tidal radius, there
would be on average more stars on circular orbits than radial orbits
at the tidal radius and the distribution would still appear hot in
projection. We explicitly checked that this is indeed the case by
considering some more extreme orbits with very large apocentres
($260$\,kpc), and pericentres large enough to make the effect of tidal
shocking negligible ($35$\,kpc). Even in these cases the projected velocity
dispersions are flat or rising, for the reasons presented above. As a
final check, we considered also only the bound stars. Plotting only
these removes the hot tidal tail, but still does not lead to cold
outer stars in projection.

We conclude that if such cold outer populations in dSph
galaxies are real then they cannot have formed as a result of tidal
effects which always produce stars which appear hot in projection.\\
\\
{\it (iii) The velocity distribution of the outer-most stars is highly
  non-Gaussian, with a significant tail of high velocity stars. This
  could lead to a large fraction of stars being deemed non-members of
  the dSph galaxy and excluded from the data analysis, leading to a
  spurious detection of a cold outer point.}\\
\\
To test this hypothesis, we plot the distribution of line of sight
velocities in a 0.2\,kpc-wide bin at a projected radius of 1\,kpc for
models B and C (Figure \ref{fig:histobin}). Over-plotted are Gaussian
fits to the binned data in both cases (red lines). The distributions
are close to Gaussian: neither shows a strong asymmetric tail. Similar
results were found at other radii. We
conclude that it is unlikely that a significant fraction of high
velocity stars would be excluded as members of Draco from an
observational data analysis.

\subsection{Strong tidal shocking: model D}\label{sec:modeld}

In model D, we consider the hypothesis that the progenitors of dSph
galaxies were much more massive than they appear at the present
time. As can be seen 
in models A, B and C, the action of tidal shocks puffs up the
satellite and lowers the velocity dispersion at all radii. In model D,
we start with a satellite which has a very large central velocity
dispersion which is not consistent with current data from the Local
Group dSphs. We consider the case where this velocity dispersion is
lowered over time by the action of tidal shocks such that it becomes
consistent with current data now.

As we showed in section \ref{sec:shocking}, it is possible to find a
regime where tidal shocks dominate over stripping interior to the
stars. The existence of such a regime is demonstrated numerically
by model D.

In Figure \ref{fig:results}, bottom panels, we can see that the
satellite has been little affected by tidal stripping inside $\sim
1$\,kpc. It retains smooth spherical contours and shows no strong
velocity gradients. Beyond $\sim 1$\,kpc, tidal stripping becomes
significant and the outer stars appear hot in projection. However, in
Figure \ref{fig:resultsden}, we can see that tidal shocking has been
much more important than the stripping over a Hubble time. The central
density of the dark matter has lowered by a factor of $\sim 10$, and
similarly for the stars. The velocity dispersion of the stars is lower
at all radii, but notice that the central value is still high -
similar to the initial conditions. This reflects the presence
initially of a central dark matter cusp. This allows the density at
the centre of the satellite to become high enough that tidal shocking
is less effective, even over very long timescales and for this extreme
orbit (recall that the satellite pericentre for model D is just
6\,kpc). Finally, notice that despite such strong tidal shocks, the
central dark matter cusp persists, albeit at a lower
normalisation. This agrees with findings from previous authors
\citep{2004ApJ...608..663K}. It
is this central dark matter cusp which leaves the satellite with a
central velocity dispersion which is too large to be consistent with
data from any of the Local Group dSphs observed so far. In a similar
model which we ran with a central dark matter core instead of a cusp,
much better agreement was obtained. We will return to this issue in
section \ref{sec:discussion}.

\begin{figure} 
\begin{center}
\epsfig{file=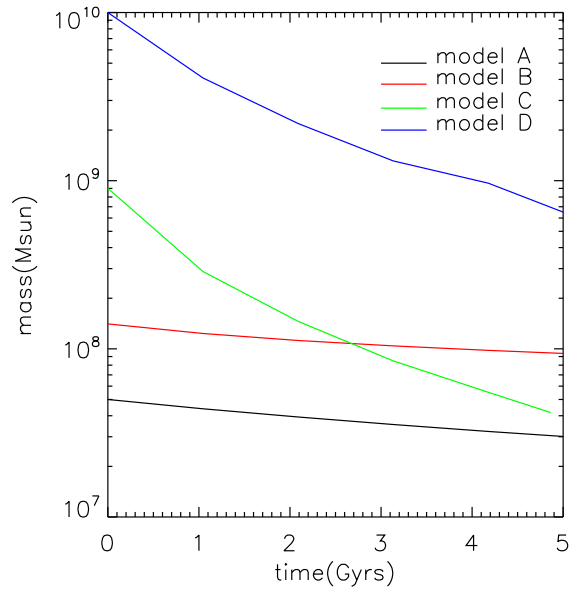,width=8.5cm}
\caption[]{Total bound mass (stars and dark matter) as a function of
  time for models A-D.} 
\label{fig:massloss}
\end{center} 
\end{figure}

In model D, as in models A and C, the satellite loses
significant mass over a Hubble time as a result of tidal stripping and
shocking. This is shown in Figure \ref{fig:massloss}, where we compare
the bound mass as a function of time for each of the models A-D. At
the output time shown, the bound mass of the satellite in model D has
been reduced by a factor of $\sim 10$. This explains why 
the final velocity dispersion matches the data from Draco, while 
the initial dispersions were much too high. Such mass loss continues
steadily over the whole simulation time due to the action of
shocks.

\section{Discussion}\label{sec:discussion}

\subsection{Sagittarius: a tidally stripped dSph galaxy}

We know of one Local Group dSph which is definitely undergoing 
tidal stripping: the Sagittarius dwarf (\bcite{2001ApJ...551..294I}
and \bcite{2005ApJ...619..807L}). Its tidal streams are quite cold,
giving a flat velocity dispersion of $\sim 10$\,km/s
\citep{2005ApJ...619..807L}, consistent with its central value
\citep{1997AJ....113..634I}. Furthermore the velocity gradients across
its minor axis ($\pm 1$\,kpc) seem to be only $\sim 3$\,km/s
\citep{1997AJ....113..634I}\footnote{The major axis results are not available
in the published literature. It is possible
that large velocity gradients may be discovered in future along the
major axis, consistent with strong tides.}.

\citet{1998ASPC..136...70O} and \citet{1997AJ....113..634I} have
argued that these data for the Sagittarius dwarf must imply large amounts
of dark matter, similar to the other dSphs, such that the effects of
tides become hard to detect within the body of the Sagittarius
dSph. An alternative view is that even strong tides do not produce
measurable effects in the kinematics, even out to radii which begin
to sample the tidal tails.  As we have shown in this paper, this
second view is possible if Sagittarius is favourably aligned (see Figure
\ref{fig:proj}). One might speculate that the other dSphs could be
undergoing tidal disruption but the signs of it are masked in
projection. Such favourable alignments seem
unlikely to be the case for all of the Local Group dSphs. Furthermore,
in at least three dSphs (UMi, Draco and Sextans), the velocity
dispersion is actually falling. This is inconsistent with strong tidal
effects interior to $\sim$\,1kpc.

\subsection{The tidal model}

\citet{1997NewA....2..139K} found that tidal models for the formation
of dSphs without dark matter can reproduce the spatial and kinematic
observations. Central to such models is the idea that the dSph galaxy
starts out as a high density star cluster. Through tidal shocking and
stripping the star cluster is heated up and becomes almost completely
unbound. In such a scenario, high projected velocity dispersions are to be
expected (c.f. model A). However, these do not correspond to a large mass
for the satellite since the tidally heated stars are not bound and the Jeans
equations or distribution function modelling do not apply.

In our simulations, we find similar results to those presented in
\citet{1997NewA....2..139K}. \citet{1997NewA....2..139K} also finds rising
(cylindrically averaged) projected velocity dispersions for 
tidally stripped and shocked dSphs (see their Figure 10). It is the
{\it new data} from the Local Group dSphs, which were not available to
\citet{1997NewA....2..139K}, that rule out these dark matter free
tidal models, not an improvement in the simulations. Even using
favourable projections (as in Figure \ref{fig:proj}), it is not
possible to have a {\it falling} projected velocity dispersion in the
tidal model - a result which concords with other studies of the tidal
model too (see e.g. \bcite{2003ApJ...592..147F}).
 
\citet{1995AJ....109.1071P} also argued against a dark matter free
tidal origin for the dSphs. They found significant velocity gradients
in their simulations which were inconsistent with observations. As in
\citet{1997NewA....2..139K}, we find that it is possible to hide such
velocity gradients using suitable projections. Thus it is the
existence of cold outer populations in projection which provide the
hardest challenge for such tidal models, rather than the velocity
gradients alone.

Finally, for Draco, \citet{2003ApJ...589..798K} showed that the
tidal model will not work since Draco cannot be very extended along
the line of sight (something which is required in any projection which
disguises the velocity gradients and significant tidal tails, but in
which the velocity dispersion is inflated due the inclusion (in
projection) of stars from the tidal tails). Through detailed modelling
of Draco, \citet{Mashchenko:2005bj} also find that the tidal model is
not favourable. Here, we confirm these results, and extend the list of
tidal stripping-free dwarfs to include UMi and Sextans.

\subsection{Mass bounds from tides}

It is interesting to now turn the problem of the cold outer stars in
Draco, Sextans and UMi on its head. The presence of such stars tells us
that either the tidal radius must lie beyond this point, or that the
cold point was formed recently and has not had time to be disrupted by
tides. The latter seems unlikely: Draco and UMi have very
old stellar populations ($\simgt 10$\,Gyrs), which suggests that any
baryonic process which might cause such a cold point must have occurred
long ago. As such, in this section,
we discuss the implications of these Local Group dSphs having tidal
radii $\simgt 0.8$\,kpc, which is the current observational edge for
kinematic observations in these galaxies. For brevity from here on, we
use the term `Local Group dSphs' to mean Draco, UMi and Sextans.

From Figures \ref{fig:results} and \ref{fig:resultsden}, we can see
that the tidal heating of stars starts at the prograde stripping
radius (this is where the tangential velocity anisotropy begins - see section
\ref{sec:stripping} and \bcite{Readinprep1}). Thus from here on, by
`tidal radius' we refer to the prograde stripping radius. 

\begin{figure} 
\begin{center}
\epsfig{file=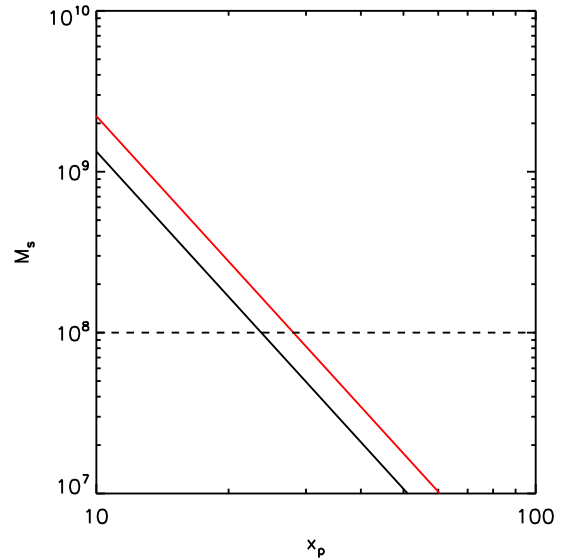,width=8cm}
\caption[]{A lower mass bound on the Local Group dSphs as a function
  of their orbital pericentres. The red line is for highly eccentric
  dSph orbits ($e=1$) while the black is for circular orbits
  ($e=0$). The mass limits were obtained from the {\it prograde}
  tidal radius (see equation \ref{eqn:rtperi} with $\alpha=1$). Also
  marked is the approximate lower bound on the mass obtained from
  stellar kinematics and distribution function models for Draco (black
  dashed line; and see \bcite{2001ApJ...563L.115K}). UMi and Sextans
  have similar lower bounds.}
\label{fig:massbound}
\end{center}
\end{figure}

The tidal radius is a function of the satellite and Milky Way
potentials, and the orbit of the satellite. However, assuming point
masses leads to, at most, errors of the order of 2-3 in the
prograde tidal radius \citep{Readinprep1}. Thus, rearranging equation
\ref{eqn:rtperi} should give us a reasonable {\it lower bound} on the
mass of Local Group dSphs {\it within the tidal radius}, as a function
of their orbital pericentre. This is plotted in Figure
\ref{fig:massbound}. It is a lower bound because the tidal radius must
be greater than the outermost measured kinematic point (in order to preserve
a cold outer point) in the Local Group dSphs, which we take to be
$\sim 0.8$\,kpc.

Since we have shown that tidal stripping is not likely to act interior
to $\sim 0.8$\,kpc 
in at least Draco, UMi and Sextans, this suggests that these galaxies
are close to equilibrium and a mass bound from the Jeans equations or
distribution function modelling is sound. What, then, can we learn
from a mass bound from tides which is degenerate with the orbit of the
dSph and the potential of the Milky Way, both of which are poorly
constrained? From Figure \ref{fig:massbound}, we can see that, if the orbital
pericentre of the Local Group dSphs is $\simgt 30$\,kpc, then we learn
nothing new from the tidal mass bound: tidal stripping and shocking
are both negligible for all satellite masses $\simgt 10^8$M$_\odot$ -
the mass bound obtained from distribution function modelling. However, if
the pericentre of the orbit is $\simlt 30$kpc, then the satellite must
have been {\it more massive in the past} than it is at present in
order to have survived intact on its current orbit. Thus for small
pericentre orbits, the tidal mass bound is a bound on the dSph's {\it
  initial} mass, before tidal stripping and shocking. That it is now
less massive than this then means that tidal shocking must have acted
to lower the central mass and velocity dispersion over a Hubble time
(as in models C and D).

\subsection{Can dSphs reside in the most massive substructure halos?}

The dark matter halo in model D had a total mass of
$10^{10}$M$_\odot$, consistent with the most
massive dark matter halos predicted by cosmological
simulations to surround the Milky Way \citep{1999ApJ...522...82K}. The
initial mass within $0.8$\,kpc for model D was 
$10^9$M$_\odot$, while after significant tidal shocking on its extreme orbit
(with pericentre of 6.5\,kpc), the final mass within $0.8$\,kpc was
$\sim 10^8$M$_\odot$. This final mass is consistent with current mass
estimates for Draco, UMi and Sextans, but its concentration is
not. The survival of the central dark matter density cusp in model D
ensured that the central velocity dispersion of the surviving dSph
galaxy was too large, even after extreme tidal shocking.

Interestingly, there may be tentative evidence that the UMi dSph
has a centrally {\it cored} dark matter density profile, rather than a
cusp \citep{2003ApJ...588L..21K}. If this is the case then it is
possible that UMi started out with total mass $10^{10}$M$_\odot$ and
had its velocity dispersion lowered through the action of tidal
shocks. With a central dark matter density core, rather than a cusp,
such shocks can lower even the central velocity dispersion to 10\,km/s
in accord with the measured values for the Local Group dSphs. Thus it
is only possible for the dSphs to inhabit the most massive
substructure halos if they have density profiles which are not
consistent with simple cosmological predictions.

A natural consequence of such strong tidal shocking is that we can expect a
correlation between the central surface brightness of the Local Group
dSphs and their pericentre: dSphs which pass closer to their host galaxies
will show lower central surface brightnesses. While the orbits of the
Local Group dSphs remain poorly constrained and so their pericentres
are not well known, a correlation between
central surface brightness and {\it distance} to the nearest host galaxy has
been observed (\bcite{1996MNRAS.278..947B} and
\bcite{alaninprep1}). While there are many possible explanations for
such a correlation (for example selection effects), a plausible
explanation would be tidal shocking. 

Finally, tidal shocks could provide the
necessary physics to remove angular momentum from the dSphs and
transform them from an initially disc-like population to a spheroidal
population as in \citet{2001ApJ...559..754M} and
\citet{2001ApJ...547L.123M}. We argue in this paper that the
Local Group dSphs have not been significantly {\it stripped} interior
to 1\,kpc, but this need not invalidate the tidal transformation model
proposed by \citet{2001ApJ...559..754M}.

\section{Conclusions}\label{sec:conclusions}
We have compared the results from a suite of N-body simulations of the tidal
stripping and shocking of two-component, spherical, dwarf galaxies in
orbit around a Milky Way like galaxy, with recently obtained kinematic
data for the Local Group dwarf spheroidals (dSphs).

Our main findings are summarised below:\\
\\
(i) Tidal stripping always leads to flat
or rising projected velocity dispersions beyond a critical radius; it
is $\sim 5$ times more likely, over random projections, that the
cylindrically averaged projected dispersion will rise than be flat.\\
\\
(ii) In several of the dSphs observed so far, there appears to be a
sharp fall-off in the projected velocity dispersion at large
radii. If such a feature is real and not a statistical artifact, it would
be rapidly destroyed by tidal stripping - tidal stripping
cannot have acted interior to the currently visible light in these
galaxies (interior to $\sim 1$\,kpc).\\
\\
(iii) By contrast, there exists a regime in
which tidal shocking can be important interior to the visible light,
even when tidal stripping is not. This could explain the
observed correlation for the Local Group dSphs between central surface
brightness and distance from the nearest large galaxy
(\bcite{1996MNRAS.278..947B} and \bcite{alaninprep1}).\\
\\
(iv) It is possible for dSphs to reside within the most massive
substructure dark matter halos ($\sim 10^{10}$M$_\odot$) and have
their velocity dispersions lowered through the action of tidal shocks,
but only if they have a central density core in their dark matter,
rather than a cusp. A central density cusp persists even after unrealistically
extreme tidal shocking and leads to central velocity dispersions which
are too high to be consistent with data from the Local Group
dSphs. dSphs can reside within cuspy dark matter halos if their halos
are less massive ($\sim 10^{9}$M$_\odot$) and therefore have smaller
central velocity dispersions initially.\\
\\
(v) A tidal origin for the formation of the Local
Group dSphs in which they contain no dark matter is strongly
disfavoured. This is because such a formation scenario requires very
strong tidal stripping, the effects of which are at odds with the
latest data from the Local Group dSphs.

\section{Acknowledgements}
JIR and MIW would like to thank PPARC for grants which have supported
this research. We would like to thank the annonymous referee for
useful comments which led to this final version.

\bibliographystyle{mn2e}
\bibliography{/home/jir22/More_space__/LaTeX/BibTeX/refs}
 
\end{document}